\documentclass[preprint,12pt]{elsarticle}
\pdfoutput=1
\usepackage[letterpaper,margin=1in]{geometry}
\usepackage{amsmath,amsthm,amssymb,commath}
\usepackage[normalem]{ulem}
\usepackage{listings,color,graphics}
\usepackage{float}
\usepackage{subfloat}
\usepackage{subfig}
\usepackage{xspace}
\usepackage{fixltx2e}
\usepackage{booktabs}
\usepackage{hyperref}

\newcommand{\USERMESO}{\texttt{\textsubscript{\textit{USER}}MESO\xspace}}

\makeatletter
\def\ps@pprintTitle{%
 \let\@oddhead\@empty
 \let\@evenhead\@empty
 \def\@oddfoot{\centerline{\thepage}}%
 \let\@evenfoot\@oddfoot}
\makeatother

\begin{document}

\begin{frontmatter}

\title{GPU-accelerated Red Blood Cells Simulations with Transport Dissipative Particle Dynamics}
\author[a]{Ansel L. Blumers}
\ead{ansel\_blumers@brown.edu}
\author[b]{Yu-Hang Tang}
\author[b]{Zhen Li}
\author[b]{Xuejin Li} 
\author[b]{George E. Karniadakis\corref{correspondingauthor}}
\cortext[correspondingauthor]{Corresponding author}
\ead{george\_karniadakis@brown.edu}

\address[a]{Department of Physics, Brown University, Providence, RI, USA}
\address[b]{Division of Applied Mathematics, Brown University, Providence, RI, USA}

\begin{abstract}
Mesoscopic numerical simulations provide a unique approach for the quantification of the chemical influences on red blood cell functionalities. The transport Dissipative Particles Dynamics (tDPD) method can lead to such effective multiscale simulations due to its ability to simultaneously capture mesoscopic advection, diffusion, and reaction. In this paper, we present a GPU-accelerated red blood cell simulation package based on a tDPD adaptation of our red blood cell model, which can correctly recover the cell membrane viscosity, elasticity, bending stiffness, and cross-membrane chemical transport. The package essentially processes all computational workloads in parallel by GPU, and it incorporates multi-stream scheduling and non-blocking MPI communications to improve inter-node scalability. Our code is validated for accuracy and compared against the CPU counterpart for speed. Strong scaling and weak scaling are also presented to characterizes scalability. We observe a speedup of $10.1$ on one GPU over all $16$ cores within a single node, and a weak scaling efficiency of $91\%$ across $256$ nodes. The program enables quick-turnaround and high-throughput numerical simulations for investigating chemical-driven red blood cell phenomena and disorders.
\end{abstract}

\begin{keyword}
dissipative particle dynamics \sep red blood cell \sep GPU \sep advection-diffusion-reaction \sep mesoscopic modeling \sep blood flow
\end{keyword}

\end{frontmatter}

\section{Introduction} \label{Introduction}

\begin{figure}[H]
	\centering
	\includegraphics[width=0.6\textwidth]{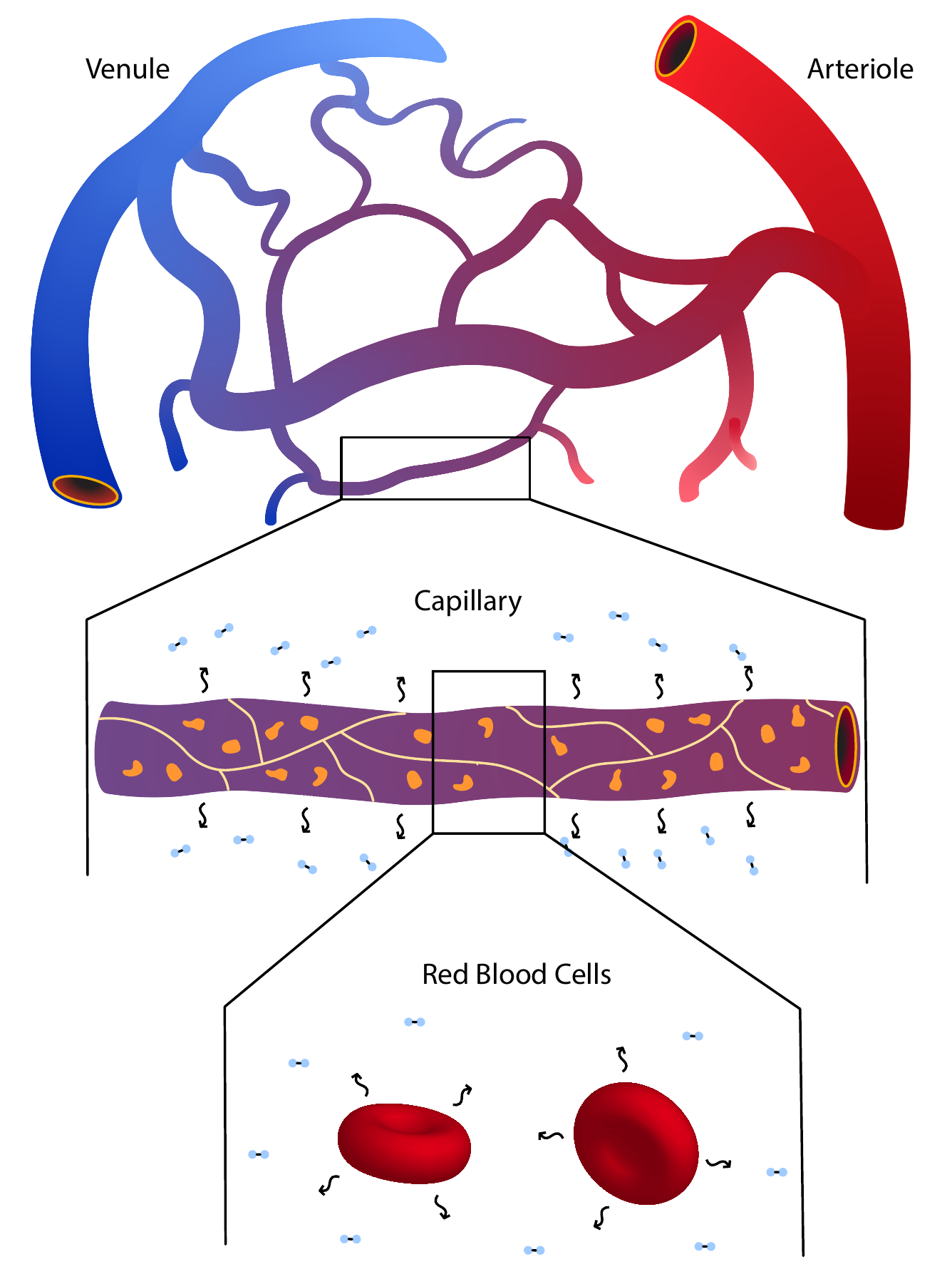}
	\caption{An illustration of oxygen release in the microvascular bed. RBCs transverse from oxygen-rich arterioles to oxygen-scarce venules through the connecting capillary beds. In each segment of the fenestrated capillary, oxygen and other nutrients diffuse into the surrounding tissue through the porous wall. The oxygen is carried and released almost exclusively by hemoglobin contained in RBCs. Our long-term goal is to simulate the chemical exchange process with explicit modeling of the RBCs as sources at the mesoscopic level. The package presented in this work serves as an enabling technology to this objective.}
	\label{fig:motivation}
\end{figure}

Blood carries nutrients, hormones, and waste products around the body. Roughly $35-45\%$ of its volume is occupied by red blood cells (RBCs) that are responsible for vital biological tasks such as circulating oxygen and carbon dioxide throughout the body. As illustrated in Figure \ref{fig:motivation}, the exchange of materials between blood and its surrounding tissues occurs primarily in the microvascular beds where arterioles and venules meet. The capillaries connecting the arterioles and venules are ideal for chemical diffusion due to their large surface-to-volume ratio and single-layered fenestrated vessel wall. However, this seemingly simple process is underpinned by intricate and detailed mechanisms. In the example of oxygen release from RBCs, the amount of oxygen discharged depends on the detailed chemical balancing between compounds such as hemoglobin and carbon dioxide, as well as ambient parameters such as plasma solubility and tissue permeability. Simulating the chemical-exchange process and capturing important details of blood flow can improve our understanding of the biological mechanisms. Such technology can be used to further our investigations into totally unexplored mechanisms linking quantitatively metabolomics with blood flow in a very precise way for the first time.

Simulations of chemical exchange in the microvascular beds with an explicit description of the source RBCs involve many biological phenomena at length scales from nanometers to micrometers. This is difficult to accomplish with either continuum descriptions based on partial differential equations or atomistic models based on classical Hamiltonian mechanics. Mesoscopic simulation methods such as Dissipative Particle Dynamics (DPD) \cite{Li2014b} are gaining momentum as a promising approach to capture these phenomena at this intermediate scale. In contract to Brownian dynamics and generalized Langevin dynamics, the pairwise force in DPD depends on the relative position and velocity between each pair of interacting particles. This ensures Gallilean invariance and momemtum conservation that allow the statistical recovery of the Navier-Stokes equation \cite{Groot1997b}. Consequently, DPD can correctly reproduce hydrodynamic behavior at the mesoscale. The versatility of DPD has been successfully demonstrated for many interesting biological applications, e.g., blood rheology \cite{Li2014c}, platelet aggregation \cite{Pivkin2009}, cell sickling \cite{Li2013}, and polymer self-assembly \cite{Li2014b,Li2015a,Tang2016}. Our newly developed transport DPD (tDPD) \cite{Li2015} can model the diffusion, advection, and reaction of the chemical transport process on top of the classic DPD framework at the molecular level using a Lagrangian framework. 

Developing a realistic RBC membrane model is crucial because the RBC mechanical and rheological state influences how a RBC performs its biological function. A coarse-grained DPD membrane model for RBC was pioneered by Pivkin \textit{et al.} \cite{Pivkin2008}. It takes bending energy, in-plane shear energy, and area and volume constraints into consideration to recover the correct bending stiffness and shear modulus. Fedosov \textit{et al.} \cite{Fedosov2010a} later extended the aforementioned model to capture the correct non-linear deformation of RBC under stress as measured from optical tweezers experiments. Although the membrane model has been carefully calibrated, large-scale RBC simulations using the DPD model remain rare while existing works are primarily limited to use a few tens of RBCs at a time. The large computational overhead has severely limited the possibility for blood-flow studies in vascular networks using the DPD-based RBC model.

The effective use of General Purpose Graphics Processing Units (GPGPUs) has improved the capability of many molecular dynamics simulation software by an order of magnitude thanks to its massively parallel nature \cite{Goetz2012,Brooks2010,Smith1996,Limbach2006,Anderson2008,Plimpton1995,Kale1999,Pronk2013}. Pushing for the peak performance on each of the 18,688 GPUs on the Titan supercomputer, Tang \textit{et al.} \cite{Rossinelli2015a} were able to simulate billions of RBCs flowing through a cancer cell filtering device. Despite the inclusion of GPUs to accelerate the simulation and to broaden the accessible time and length scale, the code was a fixed-purpose solver based on the classic DPD model. 

In this work, a generalized GPU-accelerated implementation of the RBC model with tDPD adaptation is developed to simulate RBCs with realistic physiological properties. The software features a tight integration of our earlier work on RBC membrane model, transport behavior simulations, and GPU-accelerated DPD simulators. With the new ability to track chemical concentrations, the program can be used to investigate chemical-driven or chemical-sensitive phenomena or disorders with unprecedented time and length scales. The rest of the paper is structured as follows: In Section \ref{Model}, we review the classic DPD and tDPD frameworks and present a short summary of the RBC membrane model. In Section \ref{Implementation}, we present our software implementations and algorithmic innovations. In Section \ref{Validation}, we validate the code with verification cases. In Sections \ref{Benchmark}, we demonstrate the efficiency of the code as reflected by benchmarks cases simulating pure tDPD fluids and RBC suspensions. In Section \ref{Capdemo}, we further demonstrate the capability of the software with a realistic microfluidic channel simulation. We conclude the paper in Section \ref{Summary}.

%%%%%%%%%%%%%%%%%%%%%%%%%%%%%%%%%%%%%%%%%%%%%%%%%%%%%%%%%%%%
\section{Mathematical Model}	\label{Model}     

\subsection{tDPD pairwise interaction} \label{pairwise interaction}

The classic DPD framework describes the evolution of density and velocity fields through Newton's laws. In addition to the conservation of momentum inherited from the classic DPD, tDPD also conserves the concentrations of species. Aside from its position and velocity, each particle explicitly tracts the concentrations of all species in the volume represented by this particle. A concentration gradient induces the flux of chemicals between pairs of particles, in a similar fashion to heat transfer. 

The movements of particles can be described by the following set of stochastic ordinary differential equations (ODEs) \cite{Hoogerbrugge1992}:
\begin{align}
& \frac{d\mathbf{r}_i}{dr} = v_i ,\\
& \frac{d\mathbf{v}_i}{dt} = \mathbf{F}_i = \sum_{i \neq j} (\mathbf{F}^C_{ij} + \mathbf{F}^D_{ij} + \mathbf{F}^R_{ij}),
\end{align}
where $t$, $\mathbf{r}_i$, $\mathbf{v}_i$ and $\mathbf{F}_i$ denote time, position, velocity, and force, respectively. The net force imposed on particle $i$ is the sum of conservative force $\mathbf{F}^C_{ij}$, dissipative force $\mathbf{F}^D_{ij}$, and corresponding random force $\mathbf{F}^R_{ij}$ via interactions with every particle $j$ within a radial cutoff $r_c$ of $i$. Those forces are given as \cite{Groot1997b}:
\begin{align} \label{DPDforce}
& \mathbf{F}^C_{ij} = \alpha_{ij} \; \omega_C(r_{ij}) \; \mathbf{e}_{ij}, \\
& \mathbf{F}^D_{ij} = -\gamma_{ij} \; \omega_D(r_{ij}) \; ( \mathbf{e}_{ij} \cdot \mathbf{v}_{ij} ) \; \mathbf{e}_{ij}, \\
& \mathbf{F}^R_{ij} = \sigma_{ij} \; \omega_R(r_{ij}) \; \xi_{ij} \; \Delta t^{-1/2} \; \mathbf{e}_{ij},
\end{align}
where $\mathbf{e}_{ij}=\mathbf{r}_{ij} / r_{ij}$ is the unit vector between particles $i$ and $j$. $\Delta t$ is the time step of the simulation, and $\xi$ is a symmetric Gaussian random variable with zero mean and unit variance \cite{Groot1997b}. $\alpha_{ij}$, $\gamma_{ij}$, and $\sigma_{ij}$ are conservative, dissipative, and random force coefficients, respectively. $\omega_C(r_{ij})$, $\omega_D(r_{ij})$, and $\omega_R(r_{ij})$ are the corresponding weighting functions. The relationship between dissipative and random effects is dictated by the fluctuation-dissipation theorem which imposes the following constraints \cite{Espanol1995a}:
\begin{equation} \label{FDTforce}
\sigma^2_{ij} = 2 \; k_B \; T \; \gamma_{ij}, \;\;\;\;\;\; \omega_D(r_{ij}) = \omega_R^2(r_{ij}),
\end{equation}
where $k_B$ is the Boltzmann constant, and $T$ is temperature.

The tDPD model enables us to capture reaction kinetics at the mesoscopic level. It essentially solves the advection-diffusion-reaction equation $\frac{dC}{dt} = D\nabla^2C+Q^S$, where the transport of concentration is modeled by a Fickian flux and a random flux \cite{Xu2011,Kordilla2014}. It can be described by the following equation:
\begin{equation} \label{flux}
\frac{dC_i}{dt} = Q_i = \sum_{i \neq j} ( Q^D_{ij} + Q^R_{ij} ) + Q^S_i,
\end{equation}
where $C_i$ denotes the concentration of a species carried by particle $i$, and $Q_i$ is the net flux. 
There are three flux components that can potentially alter the concentration. $Q^D_{ij}$ and $Q^R_{ij}$ denotes Fickian and random fluxes, respectively, and are given by \cite{Li2015}
\begin{align}
& Q^D_{ij} = - \kappa_{ij} \; \omega_{DC}(r_{ij}) \; (C_i - C_j), \\ 
& Q^R_{ij} = \epsilon_{ij} \; \omega_{RC}(r_{ij}) \; \Delta t^{-1/2} \; \zeta_{ij},
\end{align}
where $\kappa_{ij}$ and $\epsilon_{ij}$ adjust the strengths. The corresponding weights $\omega_{DC}(r_{ij})$ and $\omega_{RC}(r_{ij})$ are given as $\omega_{DC}=(1-r/r_{cc})^{s_{c1}}$ and $\omega_{RC}=(1-r/r_{cc})^{s_{c2}}$ with a cutoff radius $r_{cc}$, analogous to the weights in force calculations. Taking the same idea from the fluctuation-dissipation theorem in the classic DPD, the transport version requires \cite{OrtizdeZarate2006}
\begin{equation} \label{FDTconcentration}
\epsilon^2_{ij} = m^2_s \; \kappa_{ij} \; \rho \; (C_i + C_j), \;\;\;\;\;\; \omega_{DC}(r_{ij})=\omega^2_{RC}(r_{ij}),
\end{equation}
where $\rho$ is the tDPD particle density.

It is easy to see the parallel between equations (\ref{FDTforce}) and (\ref{FDTconcentration}). In particular, parameter $\kappa$ adjusts the strength of chemical transfer, as the dissipative coefficient $\gamma$ adjusts the strength of momentum dissipation. It is worth noting that each species typically has a distinct $\kappa$ value because diffusion coefficients for different chemical species are generally different. Since the mass of a single solute molecule $m_s$ is much smaller than the mass of a tDPD particle $m$ which is usually chosen to be one (normalized value), the magnitude of $\epsilon_{ij}$ is insignificant. This translates to a negligible contribution from $Q^R_{ij}$ to the diffusion coefficient. In this work,  $Q^R_{ij}$ can be safely ignored considering the simplification $m_s \ll m$ \cite{Li2015}. Finally, the source term $Q^S_i$ represents the external contributions e.g., external concentration source and boundary conditions.

The random force component in the classic DPD is stochastic by definition. The stochastic behavior of particle movement yields an additional contribution to diffusion reflected in the diffusion coefficient. This additional diffusion component $D^\zeta$ and the Fickian diffusion component $D^F$ compose the effective diffusion coefficient $D$. Because the random flux $Q^R$ is insignificant, $D^\zeta$ is due to random movement of particles solely and can be calculated by \cite{Groot1997b} 
\begin{equation}
D^{\zeta} = \frac{3k_BT}{4\pi\gamma\rho \cdot \int^{r_c}_0 r^2 \omega_D(r)g(r)dr}.
\end{equation}
The Fickian concentration flux $Q^D_{ij}$ depends on $D^F$ entirely, which can be calculated by \cite{Li2015}
\begin{equation} \label{DF}
D^F = \frac{2\pi\kappa\rho}{3} \int^{r_{cc}}_0r^4\omega_{DC}(r)g(r)dr.
\end{equation}
The physical diffusion coefficients are mapped to those of the particle simulation through parameter $\kappa$. The mapping relation can be extracted from matching simulation results with known analytical solutions as described in \cite{Li2015}.

\subsection{RBC model} \label{RBCmodel}
RBC membrane comprises a lipid bilayer supported by an inner cytoskeleton. Composed of spectrin proteins and actins in a compact network, the cytoskeleton provides structural stability. Membrane viscosity and elasticity, as well as bending stiffness, are physical properties derived from these biological components. The same physical properties can be recovered with a spring-network model that resembles a triangular mesh on a 2D surface as described in \cite{Fedosov2010a}. 

The shape of the viscoelastic membrane is maintained by a potential derived from a combination of bond, angle and dihedral interactions. The bonds, also referred to as springs, experience both an attractive and a repulsive component. The attractive potential adopts the form of the wormlike chain potential and is given by
\begin{equation}
U_{WLC} = \frac{k_BTh_m}{4p}\frac{\frac{h}{h_m}^2(3-2\frac{h}{h_m})}{1-\frac{h}{h_m}},
\end{equation}
where $k_BT$ is the energy per unit mass, $h_m$ is the maximum spring extension, and $p$ is the persistence length. The repulsive potential adopts the form of a power function given by
\begin{align}
U_{POW} = \begin{cases} 
	\frac{k_p}{(m-1)h^{m-1}}, & m \neq 1, \\
	-k_p\log(h), & m=1,
	\end{cases}
\end{align}
where $m$ is positive exponent, and $k_p$ is force coefficient. In addition to those conservative potentials, a viscous component is needed to damp the springs. It is realized by a dissipative force, as well as the corresponding random force, given as
\begin{align}
& \mathbf{F}^D_{ij} = -\gamma^T \mathbf{v}_{ij} - \gamma^C (\mathbf{v}_{ij} \cdot \mathbf{e}_{ij}) \mathbf{e}_{ij}, \\
& \mathbf{F}^R_{ij} = \sqrt{2k_BT} \Big(\sqrt{2\gamma^T} (d\mathbf{W}^S_{ij} - tr[d\mathbf{W}^S_{ij}]\mathbf{I}/3) + \frac{\sqrt{3\gamma^C-\gamma^T}}{3} tr[d\mathbf{W}_{ij}] \mathbf{I} \Big),
\end{align}
where $\gamma^T$ and $\gamma^C$ are dissipative coefficients, and $v_{ij}$ is the relative velocity. $d\mathbf{W}^S_{ij}$ is the symmetric component of a random matrix of independent Wiener increments $d\mathbf{W}_{ij}$. 

The angle interaction is facilitated by area and volume constraints given by
\begin{align}
& V_{area} = \frac{k_g(A^t-A^t_0)^2}{2A^t_0} + \sum_j \frac{k_l(A_j - A_{0,j})^2}{2A_{0,j}}, \\
& V_{volume} = \frac{k_v(V^t-V^t_0)^2}{2V^t_0},
\end{align}
where $j$ is triangle index. $k_g$, $k_l$, and $k_v$ are global area, local area, and global volume constraints coefficients, respectively. 
The instantaneous area and volume of a RBC are denoted by $A^t$ and $V^t$, whereas $A^t_0$ and $V^t_0$ represents the equilibrium area and volume. $A^t_0$ is calculated by summing the area of each triangle. $V^t_0$ is found according to scaling relationship $V^t_0/(A^t_0)^{3/2} = V^R / (A^R)^{3/2}$, where $V^R$ and $A^R$ are the experimental measurements. 

The dihedral interaction is described by the bending potential given by
\begin{equation}
V_{bending} = \sum_j k_b \Big( 1-\cos(\theta_j-\theta_0) \Big),
\end{equation}
where $k_b$ is the bending constant, and $\theta_j$ is the angle between two neighboring triangles with the common edge $j$. More detailed information regarding RBC membrane model can be found in \cite{Fedosov2010a}. 

%%%%%%%%%%%%%%%%%%%%%%%%%%%%%%%%%%%%%%%%%%%%%%%%%%%%%%
\section{Parallel Implementation} \label{Implementation}
\subsection{\USERMESO}
%- Re-introduce USERMESO

\USERMESO \cite{Tang2014} is a GPU-accelerated extension package to LAMMPS \cite{Plimpton1995} for dissipative particle dynamics and smoothed particle hydrodynamics simulations. It migrates all computation workload, except inter-rank communications, onto GPUs and achieves more than twenty times speedup on a single GPU over $8$-$16$ CPU cores. It utilizes MPI for the inter-node communication and can scale to at least $1,024$ nodes. The list below summarizes the major technical innovations involved in implementing the package:
\begin{itemize}
\item An atomics-free warp-synchronous neighbor list construction algorithm;
\item A 2-level particle reordering scheme, which aligns with the cell list lattice boundaries for generating strictly monotonic neighbor list;
\item A locally transposed neighbor list;
\item Redesigned non-branching transcendental functions ($\sin$, $\cos$, pow, $\log$, $\exp$, etc.);
\item Overlapping pairwise force evaluation with halo exchange using CUDA streams for hiding the communication and the kernel launch latency;
\item Radix sort with GPU stream support;
\item Pairwise random number generation based on per-timestep binary particle signatures and the Tiny Encryption Algorithm;
\end{itemize}

\subsection{Data layout}
Data in LAMMPS are stored in array of structure layout. To avoid strided access on the GPU, \USERMESO \; employs structure of array layout on GPU instead. A pair of interleave/deinterleave kernels facilitates the conversion between array of structure and structure of array at runtime. Meanwhile, the concentration and flux arrays of a particle must be able to hold an arbitrary number of species. Accessing this information is more efficient when the data are coalesced. Structure of Array layout is therefore chosen for storing concentrations and fluxes in CPU and Array of Structure layout in GPU.

\subsection{RBC} \label{RBCImp}

Three bonds form a triangle in the 2-D triangulated surface. Each triangle is associated with an area and a ``quasi''-volume that is calculated from the absolute distances of three vertexes. Because thread-to-thread communication is expensive, it is advantageous to compute area and volume three times, one by each thread. The excessive computation outweighs thread-to-thread communication overhead. 

The angle term demands the most computational resources because the total area and volume for each RBC need to be calculated before enforcing the area and volume constraints. This is time-consuming especially when a RBC is split between two MPI ranks. Although the areas and volumes are accumulated in a rank-basis for the portion of a RBC in that rank, the total areas and volumes must be known for each and every rank that contain any portion of that RBC. This procedure can be accomplished with an MPI-allreduce between two kernel calls for each time step. The first kernel, \textit{K-Gather}, computes the total area and volume of each RBC pertaining to that particular rank. The second kernel, \textit{K-Apply}, then enforces the area and volume constraints.  

\begin{figure}[H]
	\centering
	\subfloat[]{\includegraphics[width=0.5\textwidth]{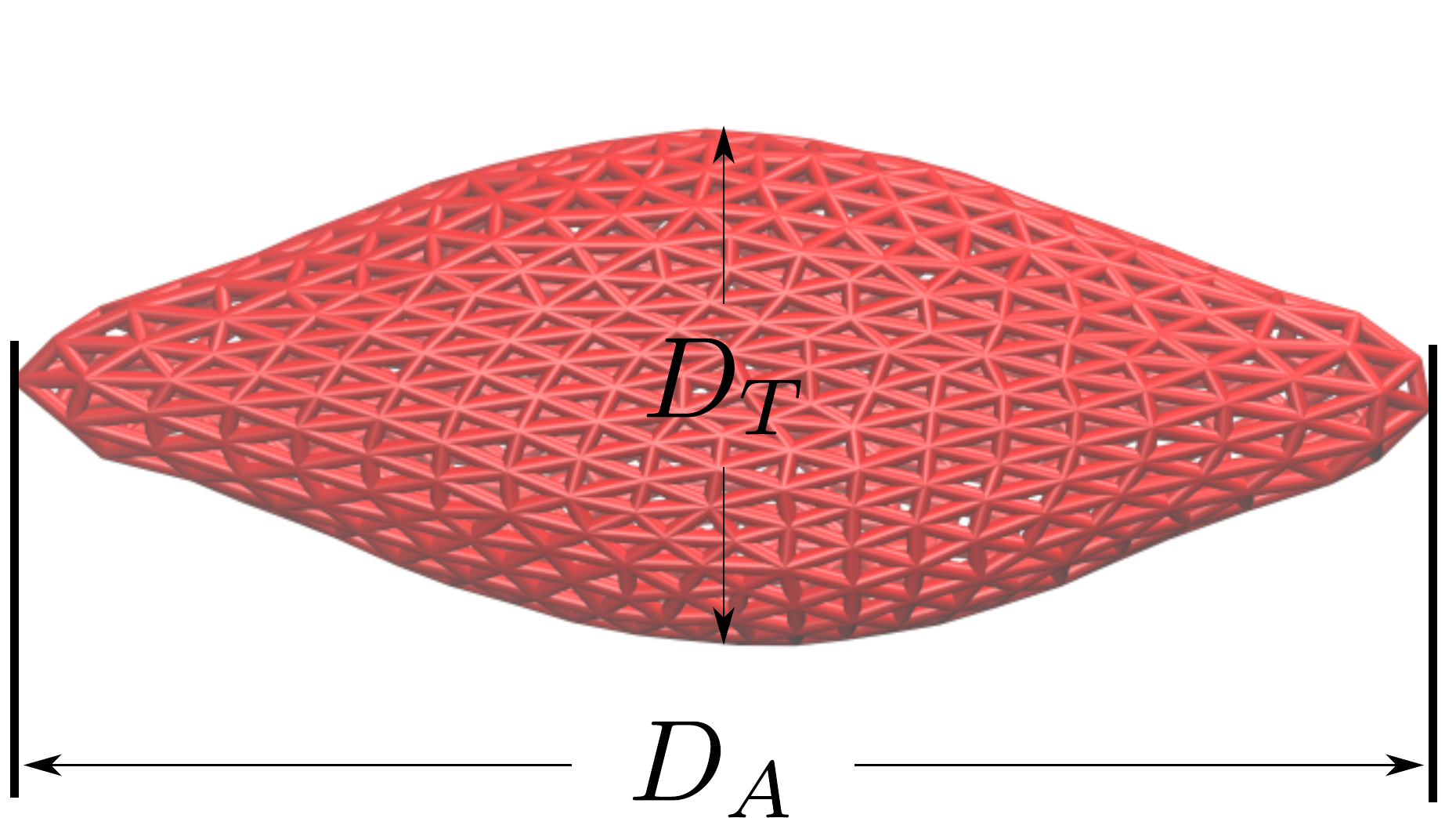}}
	\subfloat[]{\includegraphics[width=0.5\textwidth]{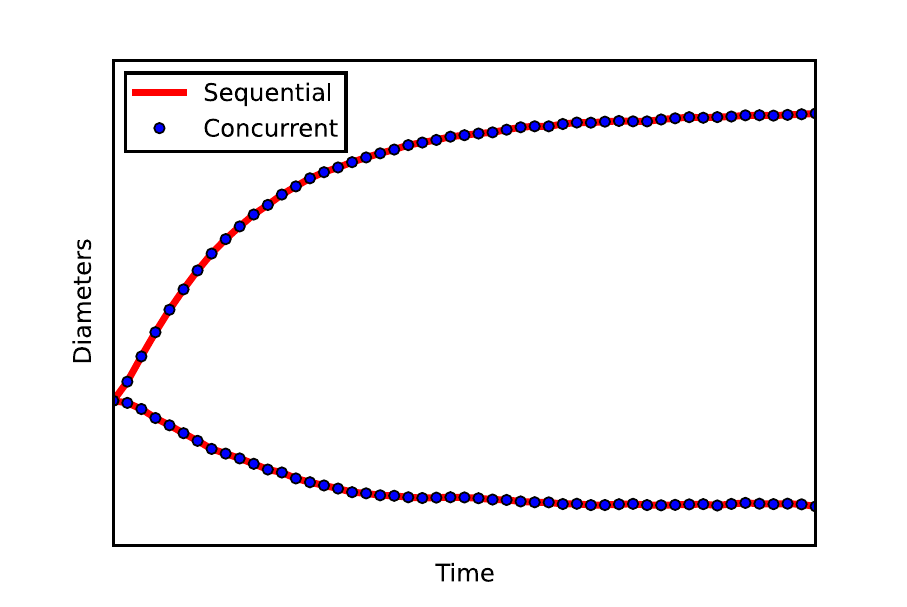}}
	\caption{A RBC stretching test was devised to show feasibility of adopting area and volume of previous time step for current time step. (a) A graphical representation of RBC stretching is shown. $D_T$ and $D_A$ denote transversal and axial diameters, respectively. (b) The time evolutions of $D_T$ (top curve) and axial diameter $D_A$ (bottom curve) are shown. The simulation results from the concurrent version matches those of the sequential version. }
	\label{fig:seqcon}
\end{figure}

This sequential procedure offers no computation overlap between those two GPU kernels because \textit{K-Apply} awaits the result of \textit{K-Gather}. However, when the time step is very small, the variation in area and volume between two consecutive steps is insignificant. Adopting area and volume from the previous time step in \textit{K-Apply} for the current time step permits concurrency between \textit{K-Apply} and MPI communication, thus overlapping CPU and GPU tasks for efficiency. A comparison with the sequential version shows visually identical results in Figure \ref{fig:seqcon}. 

\begin{figure}[H]
	\centering
	\includegraphics[width=0.5\textwidth]{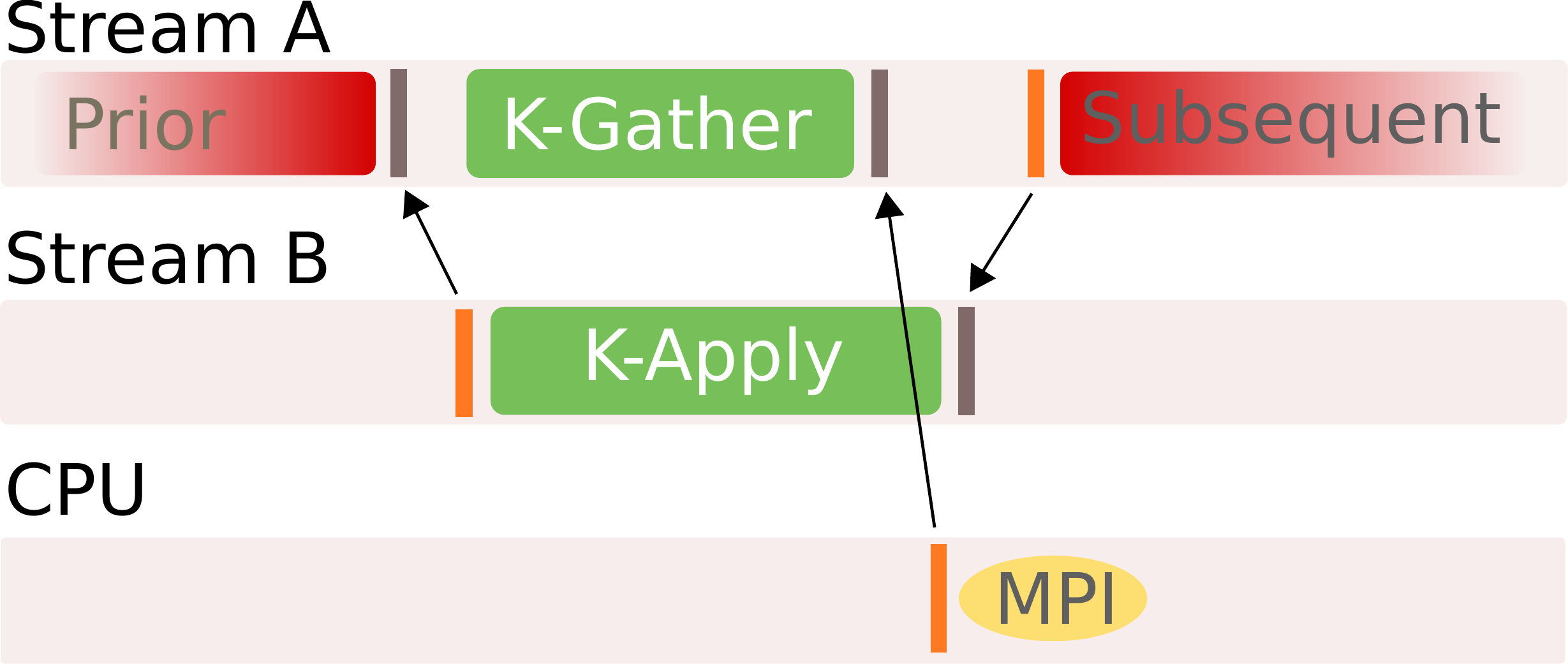}
	\caption{The multi-stream scheduling schematic drawing. The arrows denote dependency. \textit{K-Apply} has to wait until the CUDA-event marking the completion of \textit{K-Gather} is recorded. The same applies to the MPI communication.}
	\label{fig:Concurrency} 
\end{figure}

Since \textit{K-Gather} and \textit{K-Apply} are independent procedures in the concurrent version, computing \textit{K-Gather} and \textit{K-Apply} simultaneously is possible due to CUDA stream functionality. Instead of occupying the default stream exclusively, they are assigned to two streams as demonstrated in Figure \ref{fig:Concurrency}. CUDA-events are placed accordingly for synchronization. Opposite to what we expect, this setup produces worse performance. This can be explained by streaming multiprocessor saturation - two kernels competing for computing resources. The resulting higher cache refresh rate is detrimental to efficiency. Our hypothesis is validated by the fact that each kernel takes longer than its sequential counterpart to complete. 

Even though concurrent kernels setup produces undesirable performance, it prompts the possibility of simultaneous data transfer and kernel execution, as well as the possibility of non-blocking MPI communications. MPI communications are blocking by default. An imbalanced workload on each device exacerbates the latency because blocking MPI communications can only initiate when all devices synchronize with the CPUs. The more ranks there are, the bigger potential latency there is. Non-blocking MPI communication is generally preferred for this reason. In the non-blocking version of MPI-allreduce, MPI-Wait triggers Memcpy-HtD that uploads area and volume from the previous time step from CPU to GPU. Because the upload and \textit{K-Gather} execution share no dependency, Memcpy-HtD can be implemented in a different stream for efficiency. \textit{K-Apply} execution starts as soon as the result, area and volume from current time step, from \textit{K-Gather} are downloaded to CPU via Memcpy-DtH. The completion of Memcpy-DtH also triggers the non-blocking MPI communication as a concurrent process. The non-blocking MPI allows the subsequent kernel or CPU processes to continue prior to the \textit{K-Apply} completion. Communication overhead is thus greatly reduced. This setup maximizes GPU efficiency by eliminating SM un-occupancy and hiding default MPI communication overhead.

\begin{figure}[H]
	\centering
	\includegraphics[width=1.0\textwidth]{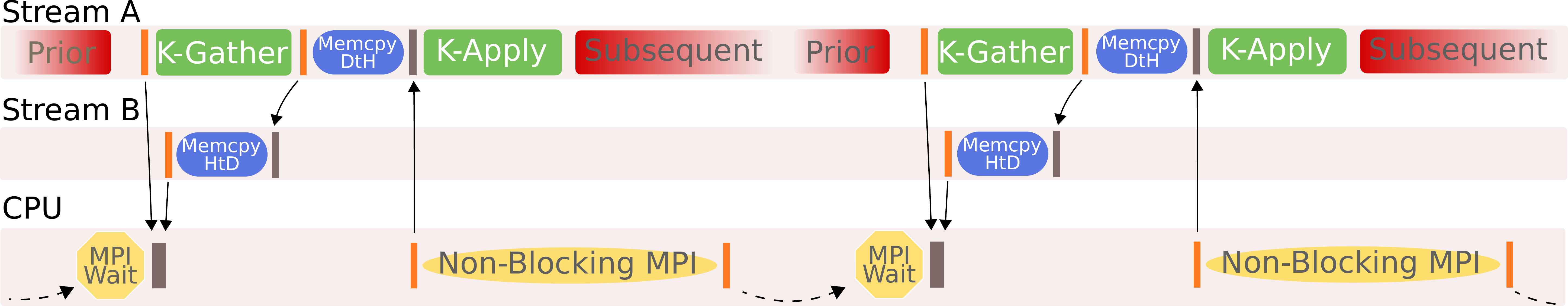}
	\caption{The solid arrows denote dependency within a time step, and the dashed arrows denote dependency across one time step. Async-Memcpys and Kernels are set up concurrently whenever possible to engage compute and copy engines simultaneously. Non-Blocking MPI eliminates the need for an additional device synchronization.}
	\label{fig:angle_nonblocking}
\end{figure}

%%%%%%%%%%%%%%%%%%%%%%%%%%%%%%%%%%%%%%%%%%%%%%%%%%%%%%
\section{Code validation} \label{Validation}
\subsection{Flow}
\begin{figure}[H]
	\centering
	\includegraphics[width=0.5\textwidth]{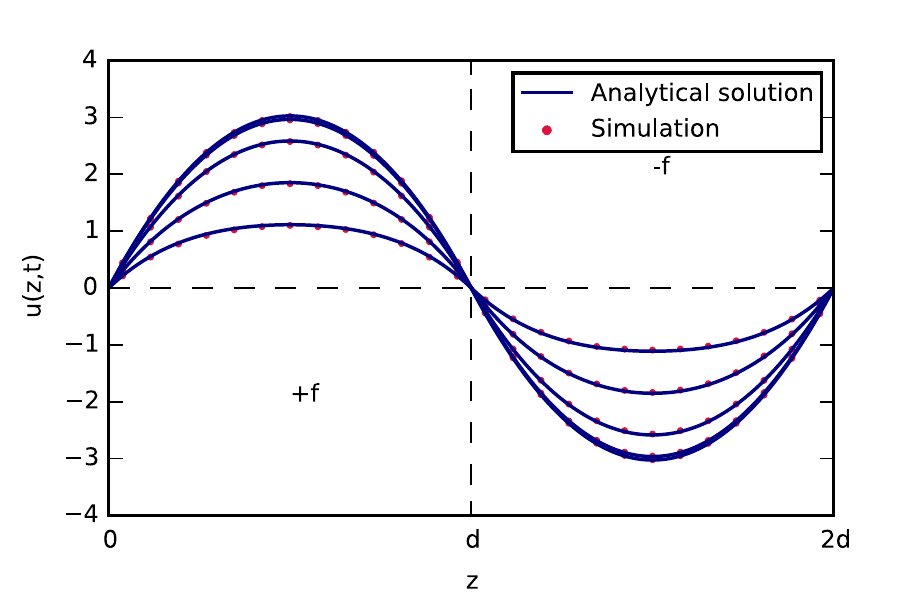}
	\caption{Transient velocity profile in a double Poiseuille flow. A body force $f$ was imposed from $0$ to $d$, and a body force $-f$ was imposed from $d$ to $2d$. The boundary in $z$ direction is thus zero by periodicity. Snapshots are taken at $t = 3, 6, 12, 24$ and $48$, corresponding to the curves from bottom to top, to show the evolution of velocity profiles.}
	\label{fig:time_evo_velocity}
\end{figure}

%- velocity double-poiseuille
Force accuracy over long-time integration is validated by simulating a transient double Poiseuille flow. The parallel flow is driven by a body force $f$ on tDPD particles. The system consists of $256,000$ particles in a $40\times40\times40$ box with periodic boundary conditions in all directions, mimicing an infinitely long channel. The simulation parameters were chosen as $\alpha=18.75$, $\gamma=4.5$, $r_c=1.58$, and $s=0.41$. Results shown in Figure \ref{fig:time_evo_velocity} are in good agreement with the analytical solution in \cite{Sigalotti2003}.

%%%%%%%%%%%%%%%%%%%%%%%%%%%%%%%%%%%%%%%%%%%%%%%%%%%%%
\subsection{Transport}
The diffusive property is validated by solving a one-dimensional diffusion equation in a infinitely long cylinder with circular cross-section. The cylinder with radius $R=20$ is composed of $402,073$ particles in a $80\times40\times40$ domain. The simulation parameters associated with force calculation were chosen as $\alpha=18.75$, $\gamma=4.5$, $r_c=1.58$, and $s=0.41$. For flux calculations, the simulation parameters were chosen as $s_2=2.0$, $\kappa=5.0$, and $r_{cc}=1.58$ as explained in Section \ref{pairwise interaction}. An initial uniform concentration $C_0$ and constant Dirichlet boundary $C_D$ were implemented. The Dirichlet boundary condition is realized with the effective boundary method demonstrated in \cite{Li2015}. This method also satisfies the no-slip boundary condition \cite{Li2013,Li2014}.

\begin{figure}[H]
	\centering
	\includegraphics[width=0.5\textwidth]{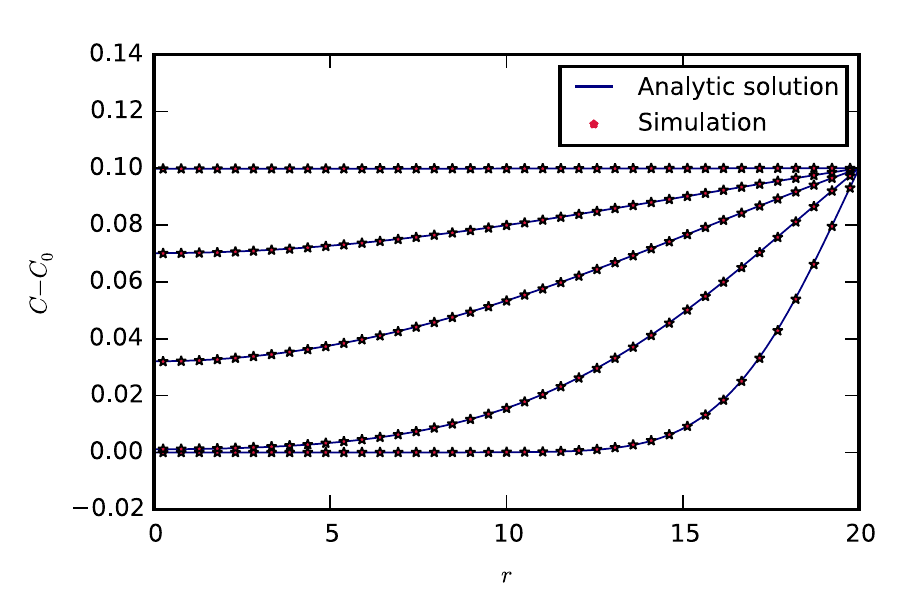}
	\caption{ Time evolution of concentration profiles in an axial symmetric infinitely long tube. Initially the system has concentration $C_0$ everywhere. The boundary is kept at $C_0+0.1$ for $t>0$. Snapshots are taken at $t = 1, 5, 15, 30$ and $100$ to show the evolution, from the bottom curve to the top. }
	\label{fig:time_exact_dirichlet}
\end{figure}

The exact solution $C(t)$ to the diffusion equation $\frac{\partial C}{\partial t}= \frac{1}{r} \frac{\partial}{\partial r}( rD\frac{\partial C}{\partial r})$ is given as \cite{Crank1980}
\begin{equation} \label{eq:DirichletDiffusion}
\frac{C(t)-C_0}{C_D-C_0} = 1 - \frac{2}{R} \sum^\infty \frac{J_0(\lambda_n r)}{\lambda J_1(\lambda_n R)} \cdot \exp(-D\lambda^2 t),
\end{equation}
where $R$ is the radius, and $\lambda_n$ are the positive roots of Bessel functions of the first kind $J_0(R\lambda_n)=0$. The simulation result matches the analytical solution as shown in Figure \ref{fig:time_exact_dirichlet}.

%%%%%%%%%%%%%%%%%%%%%%%%%%%%%%%%%%%%%%%%%%%%%%%%%%%%%
\subsection{RBC elasticity}
\begin{figure}[H]
	\centering
	\includegraphics[width=0.5\textwidth]{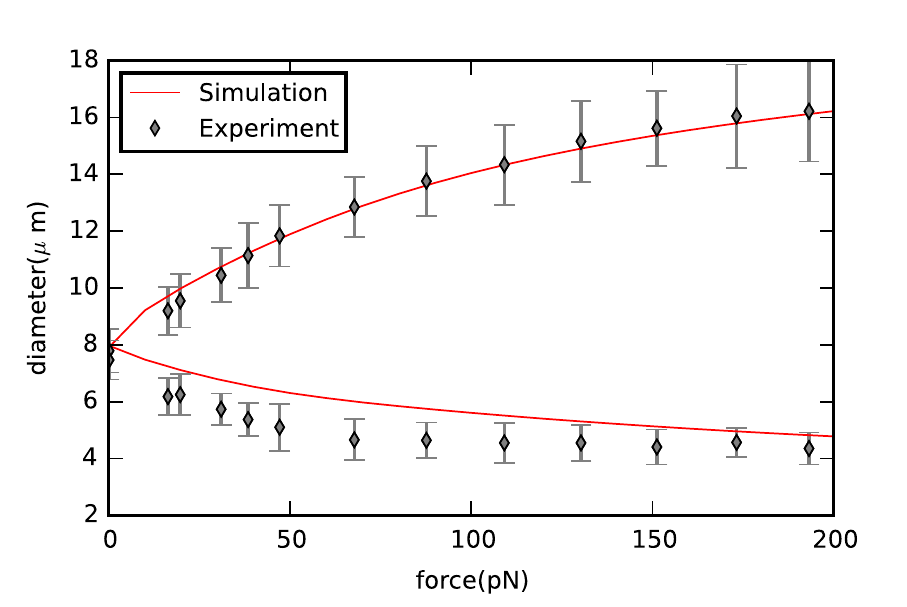}
	\caption{RBC elasticity validation. The top curve represents axial diameter $D_A$, and the bottom curve represents transverse diameter $D_T$. As Fedosov \textit{et al.} \cite{Fedosov2010a} pointed out, the small discrepancy in transverse diameter between simulation and experiment is probably due to the optical measurement being performed from only one angle.}
	\label{fig:opttweezer}
\end{figure}

For validation, the RBC implementation was compared with experimental data \cite{Suresh2015} from a RBC stretching experiment via optical tweezers. The stretching was simulated by applying a constant force on few particles in opposing ends of a RBC \cite{Fedosov2010a}. By varying the stretching force, the axial diameter $D_A$ and transverse diameter $D_T$ adjust according to the membrane elasticity. Excellent matching between the simulation and the experiment was obtained, as shown in Figure \ref{fig:opttweezer}. The result is identical to the one obtained using a different code by Fedosov \textit{et al.} \cite{Fedosov2010a}, where there is also an explanation of the disparity with the experiments of the lower branch in the plot. 

%%%%%%%%%%%%%%%%%%%%%%%%%%%%%%%%%%%%%%%%%%%%%%%%%%%%%%
\section{Benchmark} \label{Benchmark}
\subsection{Method}
The benchmarks were run on Titan at the Oak Ridge National Laboratory. Titan is a Cray XK7 system employing 18,688 nodes, each containing an AMD $Opteron$ 6274 CPU and a NVIDIA $Tesla$ K20X GPU. Every CPU contains 16 integer cores and 8 floating point cores clocked at 2.2GHz. The Kepler architecture GPU has 2,688 CUDA cores with a peak double-precision performance of 1.31 TFLOPS. The CPU solver is compiled with GCC, $-O3$ optimization. The GPU version uses NVIDIA NVCC compiler with $-O3$ optimization. 

Two system setups, pure tDPD fluid and RBC suspension, were tested. The pure tDPD fluid system consists of simple tDPD particles with pair-interaction only, whereas the system of RBC suspension contains a combination of pure tDPD fluid and coarse-grained RBCs. For a fair comparison, only the main execution loop was timed.

\subsection{Single node speedup} 

\begin{figure}[H]
	\centering
	\includegraphics[width=0.5\textwidth]{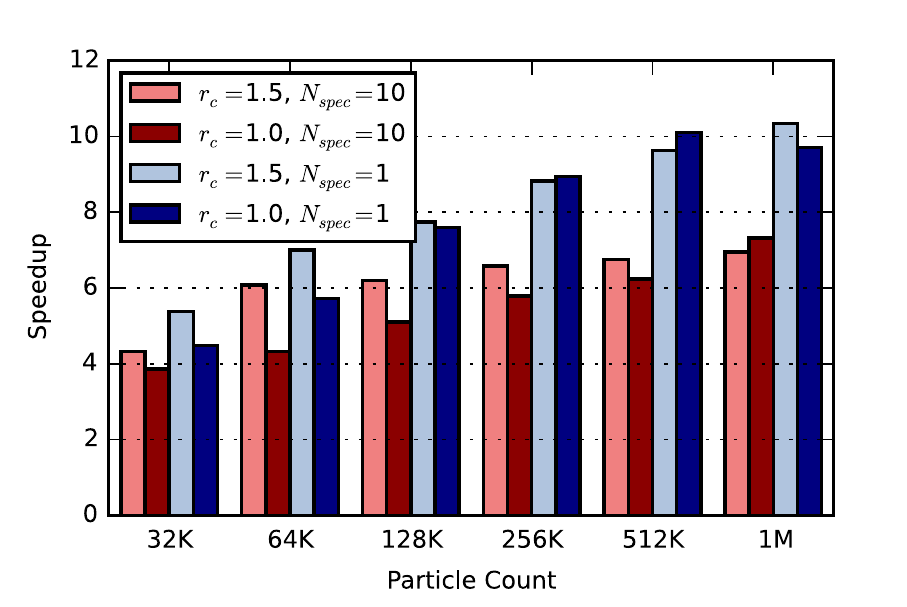}
	\caption{Speedup of pure tDPD fluid simulation over the CPU solver. Two different neighbor list cutoff values, $r_c=1.0$ and $1.5$, as well as two different numbers of chemical species, $N_{spec}=1$ and $10$, were considered. Neighbor lists were updated every $5$ steps.}
	\label{fig:Fluid_Single_Speedup}
\end{figure}

\paragraph{Pure tDPD fluid}
The key performance metric used in this work is speedup, defined as a the ratio of time elapsed between the CPU and GPU implementations. The benchmark reveals significant speedup up to $10.3$ times over the CPU solver for different parameter values as shown in Figure \ref{fig:Fluid_Single_Speedup}. When a large number of species is included, the performance declines because the computation becomes memory-bound. Although $N_{spec}$ is an impacting factor, the GPU version still delivers up to $7.2$ times speedup even at $N_{spec}=10$. Solutions such as reduced precision or data compression are viable to alleviate the bottleneck.

\paragraph{RBC suspension}
Systems with two different Hematocrit ($Hct$), $7$ and $35$, were benchmarked. Each RBC is represented by $500$ tDPD particles. The number of total tDPD particles is the sum of tDPD solvent particles and RBC particles (Table \ref{table:conversion}). In this case, only one species was included. The speedup is comparable to pure tDPD fluid of similar total particle count.

\begin{table}[H]
	\centering
	\small 
	\begin{tabular}{ l  l  l  l  l  l l l }
		\toprule		
		Hct & System & Solvent & RBC & Total Particle & Speedup\\
		& Volume & Particles & Count & Count & \\
		\midrule
		$7\%$ & $8,192$  & $32,768$ & $6$ & $35,768$ & $3.8$\\
		 & $16,384$ & $65,536$ & $12$ & $71,536$ & $5.1$\\
		 & $32,768$ & $131,072$ & $24$ & $143,072$ & $5.4$\\
		 & $65,536$ & $262,144$ & $49$ & $286,644$ & $5.7$\\
		$35\%$ & $8,192$ & $32,768$ & $30$ & $47,768$ & $4.5$\\
		 & $16,384$ & $65,536$ & $61$ & $96,036$ & $5.3$\\
		 & $32,768$ & $131,072$ & $123$ & $192,572$ & $5.9$\\
		 & $65,536$ & $262,144$ & $246$ & $385,144$ & $6.7$\\
		\bottomrule	
    \end{tabular}    
    \caption{ Speedup of RBC suspension simulation over the CPU solver. The particle count in a domain depends on $Hct$. For example, $Hct=7$ for the system volume of $8,192$ translates to six RBCs. Combining with $32,768$ pure fluid particles, this RBCs in fluid system has $35,768$ tDPD particles total.}
    \label{table:conversion}
\end{table}

We also tested the performance of the code on newer GPU architectures, namely Maxwell and Pascal (Table \ref{table:newGPUs}). The workstation used for benchmark is equipped with two Intel Xeon E5-2630L CPUs at 2.0 GHz, one GeForce TITAN X Maxwell GPU, and one GeForce TITAN X Pascal GPU. The speedup is measured as the ratio between the simulation wall time with 1 GPU or 8 CPU cores in the workstation. 

\begin{table}[H]
	\centering
	\small
	\begin{tabular} { l l l l }
		\toprule
		System & Total Particle & TITAN X Maxwell& TITAN X Pascal\\
		Volume & Count & Speedup & Speedup \\
		\midrule
		$8,192$ & $47,768$ & $4.8$ & $6.5$ \\
		$16,384$ & $96,036$ & $5.8$ & $9.2$ \\
		$32,768$ & $192,572$ & $7.2$ & $9.9$ \\
		$65,536$ & $385,144$ & $7.2$ & $10.1$ \\
		\bottomrule
	\end{tabular}
	\caption{Speedup of RBC suspension simulation on TITAN X Maxwell and Pascal over the CPU solver. Under the same simulation domain setup with Hct $35\%$, the GPUs with newer architectures outperform the K20X (Kepler) that is currently employed at Oak Ridge National Laboratory. The larger speedup is contributed by the increased bandwidth and higher top speed.}	
	\label{table:newGPUs}
\end{table}

%%%%%%%%%%%%%%%%%%%%%%%%%%%%%%%%%%%%%%%%%%%%%%%%%%%%%%%
\subsection{Weak \& Strong Scalings}
% write about tDPD fluid simulation setup and RBC simulation setup
The same neighbor list update frequency and $r_c$ from the single node benchmarks were used to establish the weak scaling and strong scaling benchmarks. The metric (million particles) $\cdot$ (steps per second), or \textit{MPS/second}, is used for absolute performance characterization and comparison across different systems.

\begin{figure}[H]
	\centering
	\includegraphics[width=0.5\textwidth]{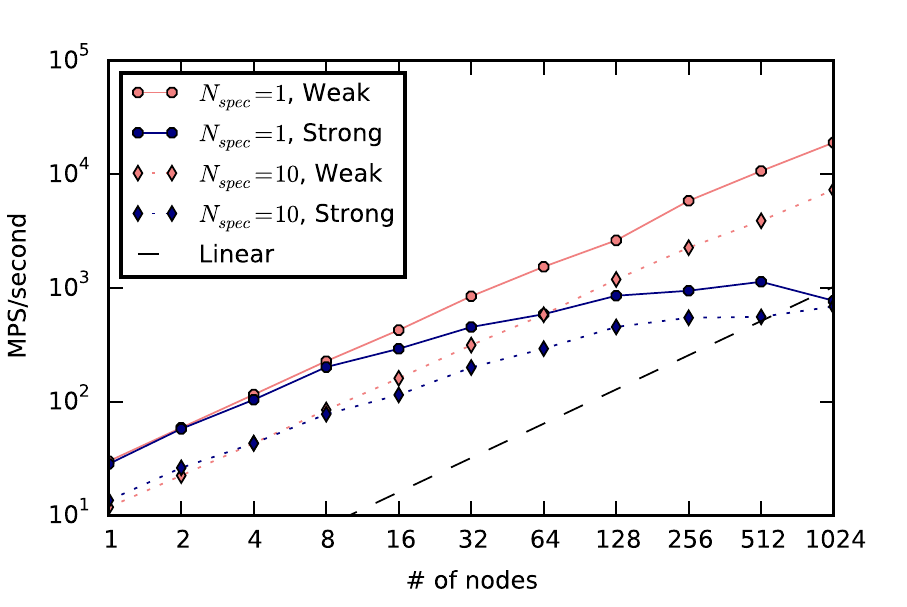}
	\caption{Weak scaling and strong scaling of pure tDPD fluid particles in a log-log plot.}
	\label{fig:Weak_Strong_fluid}
\end{figure}

\paragraph{Pure tDPD fluid}
Weak and strong scalings were benchmarked on up to $1,024$ nodes. With $524,288$ particles per node, weak scaling demonstrates nearly linear scaling. Strong scaling with a fixed system size of $2,097,152$ particles is also presented in Figure \ref{fig:Weak_Strong_fluid}. Absolute performance plateaus around $512$ nodes, where the parallel overhead cancels any extra computational resources. At $1,024$ nodes, the parallel overhead and data transfer completely dominate the computation.

\begin{figure}[H]
	\centering
	\begin{tabular}{c c}	
	\subfloat[]{\includegraphics[width=0.5\textwidth]{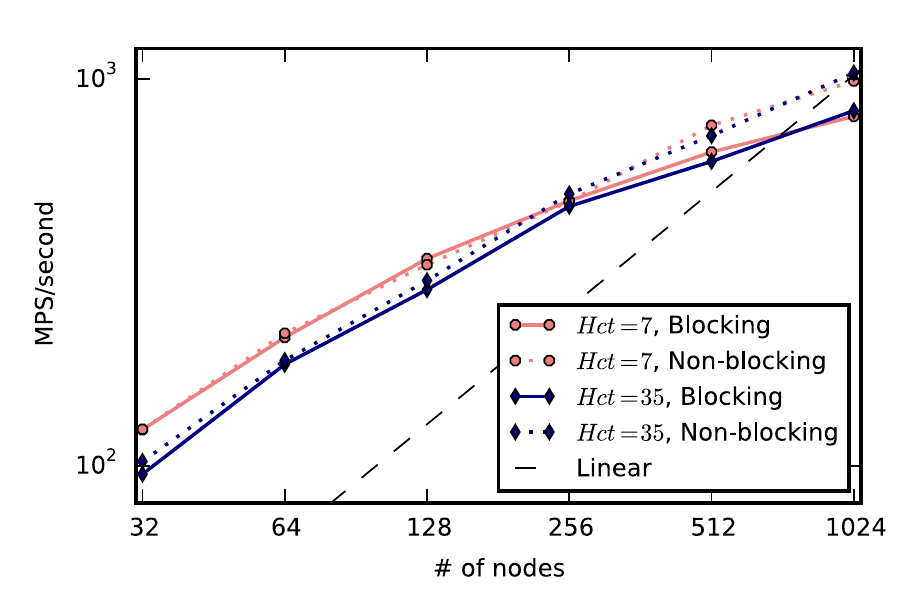}} &
	\subfloat[]{\includegraphics[width=0.5\textwidth]{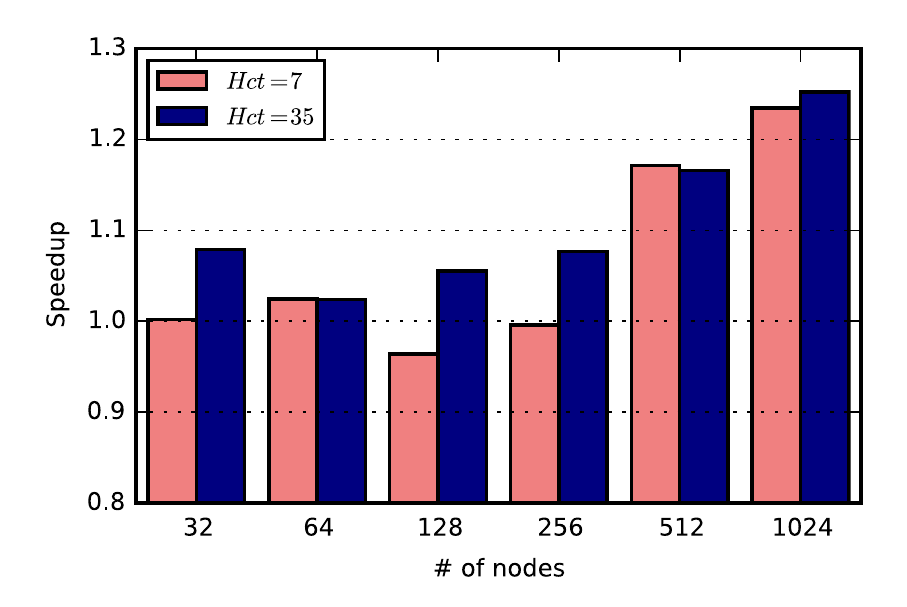}} \\
	\subfloat[]{\includegraphics[width=0.5\textwidth]{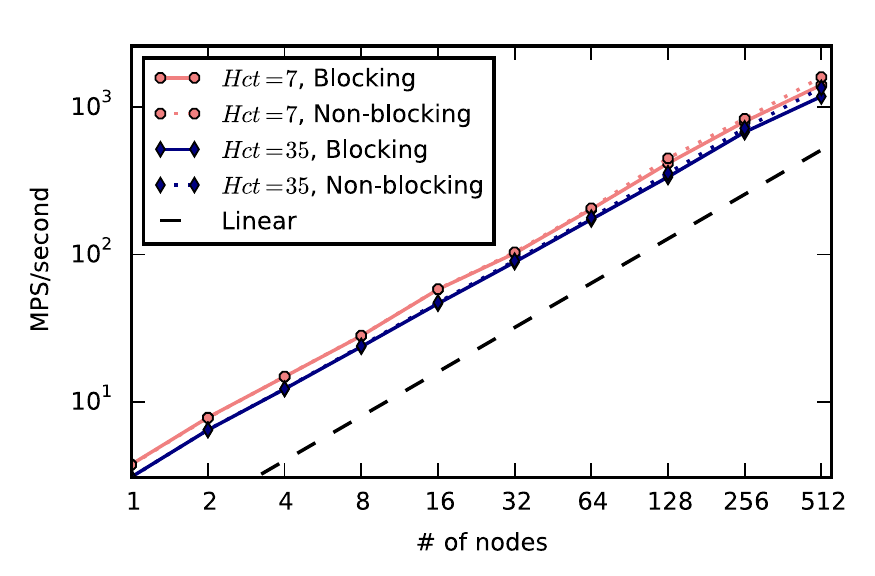}} &
	\subfloat[]{\includegraphics[width=0.5\textwidth]{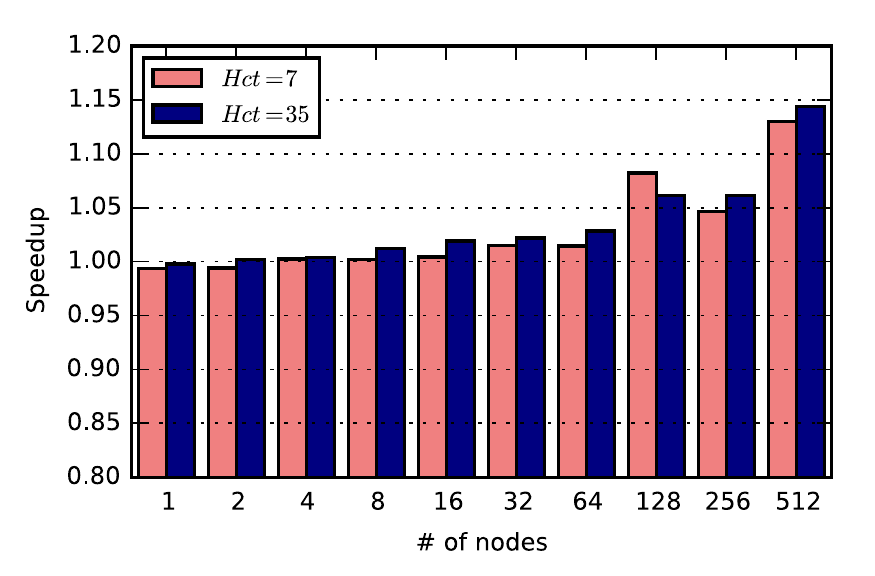}} \\
	\end{tabular}
	\caption{ Log-log plots of (a) strong scaling and (c) weak scaling of the RBC suspension system. Compared to the blocking versions, the non-blocking counterparts of (b) strong and (d) weak scaling improve scalability at high node counts.}
	\label{fig:RBCStrongWeakSU}
\end{figure}

\paragraph{RBC suspension}
Strong scaling with a system volume of $2,097,152$ is shown in Figure \ref{fig:RBCStrongWeakSU}. It is clearly seen that the non-blocking implementation of the angle term is more efficient at large node counts. The slowly diverging absolute performance plots reveal the increasing penalty caused by blocking MPI communications. Switching to non-blocking reduces the execution time by approximately $25\%$ in the case of $1,024$ nodes. Weak scaling of system volume of $32,768$ per node also shows nearly linear scalability. The non-blocking version clearly delivers a better scalability. 

%%%%%%%%%%%%%%%%%%%%%%%%%%%%%%%%%%%%%%%%%%%%%%%%%%%%%%
\section{Capability demonstration}\label{Capdemo}
%- Arbitrary Boundary Condition / RBCs
The role of RBCs in microcirculation has been investigated using microfluidic devices \cite{Holm2011,Jaggi2007,Bhagat2011}. However, certain properties such as chemical concentrations are difficult to capture experimentally with high spatial and temporal resolution. The technology presented in this work can capture the concentration fields of an arbitrary number of species in one simulation.

\begin{figure}[H]
	\centering
	\includegraphics[width=0.5\textwidth]{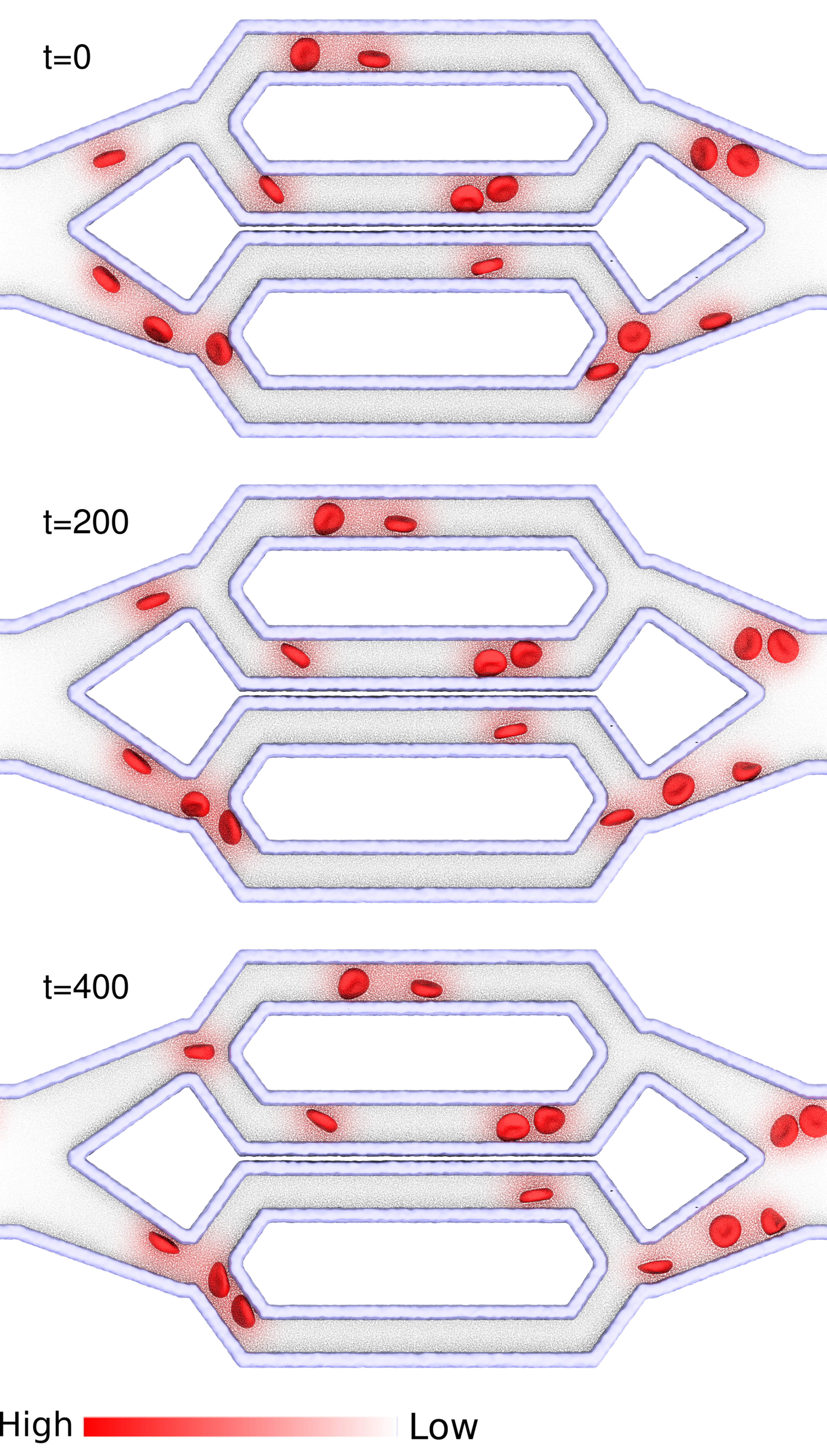}
	\caption{An example simulation combining RBC model and tDPD formulation is shown here. The RBC-fluid mixture flows through a microfluidic device from left to right. The white-red intensity field represents concentration gradient qualitatively. The chemical released by RBCs diffuses and dissipates in the fluid. The snapshots were taken at time $0$, $200$, and $400$.}
	\label{fig:showcase}
\end{figure}

To demonstrate its versatility, chemical diffusion from a RBC suspension flowing in a microfluidic device was simulated. As the chemical diffuses from the RBCs, it reacts with the solvent and dissipates. In reality, the dissipation can be induced by a number of causes, such as chemical reaction, disintegration and evaporation. The white-red intensity field shown in Figure \ref{fig:showcase} represents concentration gradient qualitatively. The redish tone surrounding RBCs indicates abundant presence of the chemical near the sources. 

Among the $720,778$ tDPD particles in the simulation, roughly $95\%$ represent the fluid and the solid channel wall at a density of $4$ particles/volume. For each RBC, the membrane is composed of $500$ particles connected by a network of springs, and the cytoplasm is portrayed by $372$ particles enclosed by the membrane. The cytoplasm and fluid particles reside in their respective sides, enforced by a novel boundary resolving technique \cite{Li2017}. This technique is also applied on the fluid-wall interface to prevent penetration of fluid particles. The same simulation setup will take approximately $12$ times as long on the CPU, deduced from the benchmark results. 

%%%%%%%%%%%%%%%%%%%%%%%%%%%%%%%%%%%%%%%%%%%%%%%%%%%%%%
\section{Summary}  \label{Summary}
In this paper, a GPU-accelerated RBC simulation package based on a tDPD adaptation of our RBC model \cite{Fedosov2010a} is presented. Taking advantage of the ability to model chemical transport in tDPD, RBC dynamics and the advection-diffusion-reaction processes can be simulated simultaneously. The effective use of GPUs improves the code performance dramatically as revealed by the benchmarks done on Titan at Oak Ridge National Lab. With some novel algorithms in streamlining the RBC computation, our code was able to produce up to $10.1$ times speedup when compared with its CPU counterpart on a single-node. The weak scaling benchmarks show almost linear scaling, while further speedup is possible even beyond $1,024$ nodes as indicated by strong scaling benchmarks. Furthermore, we demonstrated the software's capability by simulating chemical diffusion in a RBC suspension traversing inside a microfluidic device. Incorporating the boundary resolving technique that deals with arbitrary shapes \cite{Li2017}, the software can easily reconstruct complex experimental apparatuses and perform realistic RBC simulations.

It should be stressed that the software presented in this study is highly customizable, as opposed to the fixed-purpose program by Rossinelli \textit{et al.} \cite{Rossinelli2015a}. We encourages researchers to adopt our user-friendly software in their research. The software is freely available on Github, following the link \url{https://github.com/AnselGitAccount/USERMESO-2.0}

%%%%%%%%%%%%%%%%%%%%%%%%%%%%%%%%%%%%%%%%%%%%%%%%%%%%%%
\section*{Acknowledgments}
This work was supported by NIH grants U01HL114476 and U01HL116323. The benchmarks were performed on Titan at Oak Ridge National Laboratory through the Innovative and Novel Computational Impact on Theory and Experiment program under project BIP118. A.L.B. would like to acknowledge Wayne Joubert (Oak Ridge National Laboratory) for his effort to coordinate machine reservations. We also gratefully acknowledge the support from Mark Berger of NVIDIA Corporation with the donation of the GeForce TITAN X Pascal GPU used for this research.

%%%%%%%%%%%%%%%%%%%%%%%%%%%%%%%%%%%%%%%%%%%%%%%%%%%%%%
\setcounter{figure}{0}
\renewcommand{\thefigure}{A\arabic{figure}}
\section*{Appendix}
\begin{figure}[H]
	\centering
	\includegraphics[width=0.5\textwidth]{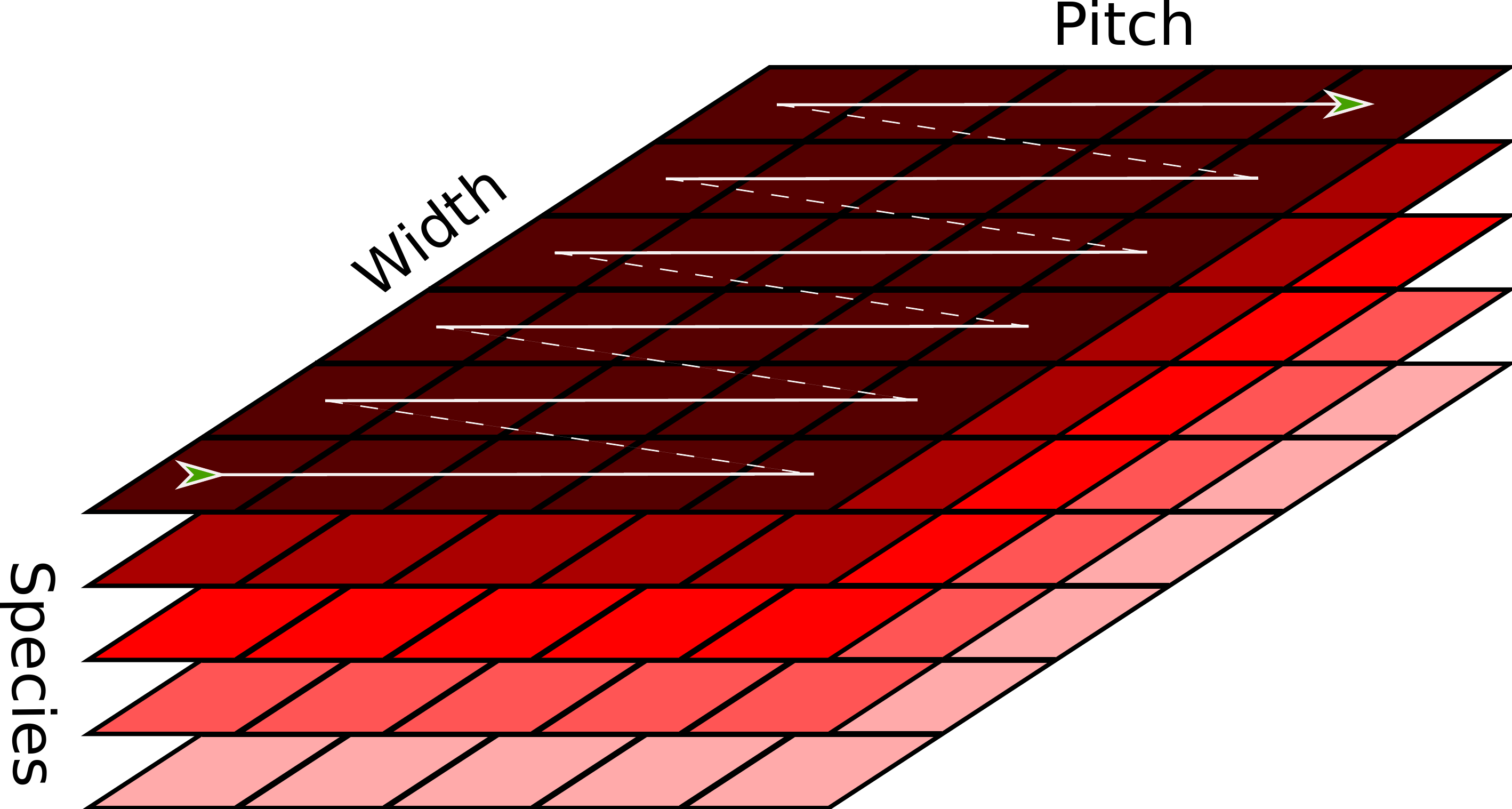}
	\caption{Schematic representation of 2D layered texture. Each layer holds the concentration of all particles for a particular species. The particles then wrap around to form a 2D array within each layer.}
	\label{fig:2DLayTex}
\end{figure}

Another attempt in speeding up the program was to reduce concentration access-time on the GPU. Layered textures are commonly used due to their optimization on accessing data with spatial locality. Because 1D layered texture has a valid extent upper limit of $16,384$ for Maxwell GPU architecture, 2D layered texture illustrated in Figure \ref{fig:2DLayTex} must be used to hold a reasonable number of particles. Unexpectedly, the layered texture memory implementation performs worse than the non-texture version revealed by benchmarks. Nvidia GPU profiling pinpoints ``data request" as the main stall reason. The much diminished texture hit rate indicates that the limited texture cache resources are experiencing cache depletion. The data locality in coordinates and velocity texture cache are strategically optimized for cache hit rate. The concentration texture data deplete the cache and thus disrupt the data locality optimization. The overall texture cache hit rate is therefore reduced significantly, and this translates directly to worse performance. 

%%%%%%%%%%%%%%%%%%%%%%%%%%%%%%%%%%%%%%%%%%%%%%%%%%%%%%
\section*{Reference}
\bibliographystyle{cpc}
\bibliography{arXiv_submission_Nov162016.bib}

\end{document}